\documentstyle[aps,preprint]{revtex}
\begin{document}
\draft
\preprint{\today,USC(NT)-97-01, nucl-th/96xxxxx}
\title{Chiral Perturbation Theory and
the $pp \rightarrow pp\pi^0$ Reaction \\
Near Threshold}
\author{T. Sato,$^a$ T.-S. H. Lee,$^b$ F. Myhrer$^c$
and K. Kubodera$^c$}

\address{$^a$Department of Physics, Osaka University,
Toyonaka, Osaka 560, Japan}
\address{$^b$Physics Division, Argonne National Laboratory,
Argonne, Illinois 60439-4843, U. S. A.}
\address{$^c$Department of
Physics and Astronomy,
University of South Carolina \\
Columbia, SC 29208, U. S. A.}


\maketitle
\vspace{-0.5cm}
\begin{abstract}
A chiral-perturbative consideration
of the near-threshold
$pp \rightarrow pp\pi^0$ reaction indicates
that the pion-rescattering term
has a substantial energy and momentum dependence.
The existing calculations
that incorporate this dependence
give pion rescattering contributions
significantly larger than those of
the conventional treatment, and
this enhanced rescattering term interferes destructively
with the one-body impulse term,
leading to theoretical cross sections that are
much smaller than the observed values.
However, since the existing calculations are
based on coordinate-space representation, they
involve a number of simplifying assumptions
about the energy-momentum flow
in the rescattering diagram,
even though the delicate interplay between
the one-body and two-body terms makes it desirable
to avoid these kinematical assumptions.
We carry out here a momentum-space calculation
that retains the energy-momentum dependence
of the vertices as predicted by chiral perturbation theory.
Our improved treatment increases
the rescattering amplitude
by a factor of $\sim$3 over the value obtained
in the r-space calculations.
The $pp \rightarrow pp\pi^0$ transition amplitude,
which is now dominated by the rescattering term,
leads to the cross section much larger than what was reported
in the approximate r-space calculations.
Thus, the extremely small cross sections
obtained in the previous chiral perturbative treatments of
this reaction should be considered as an accidental consequence
of the approximations employed rather
than a general feature.

\bigskip
\bigskip


\end{abstract}

\narrowtext
\section{Introduction}

The high-precision measurements\cite{meyetal90,Uppsala}
of the total cross sections near threshold
for the $pp \rightarrow pp\pi^0$ reaction
have invited many theoretical investigations
on this process \cite{ms91}-\cite{kmr96}.
The pion production near threshold is expected to occur via
the single-nucleon process (the impulse or Born term), Fig.1(a),
and the $s$-wave pion rescattering process, Fig.1(b).
In the conventional treatment\cite{kr66},
the $\pi$-$N$ vertex for the impulse term is
assumed to be given by the Hamiltonian
\begin{equation}
{\cal H}_0 = \frac{g_A}{2 f_\pi} \bar{\psi}
\left( \vec{\sigma} \!\cdot\!
\vec{\nabla}
(\mbox{\boldmath$\tau$} \!\cdot\!
\mbox{\boldmath$\pi$} )
- \frac{i}{2m_{\mbox{\tiny N}}}
\{ \vec{\sigma} \!\cdot\!
\vec{\nabla},
\mbox{\boldmath$\tau$} \!\cdot\!
\dot{\mbox{\boldmath$\pi$}} \}
\right) \psi,
\label{eq:H0}
\end{equation}
where $g_A$ is the
axial coupling constant,
and $f_\pi$ = 93 MeV is the pion decay constant.
The first term gives
$p$-wave pion-nucleon coupling,
while the second term accounts
for the nucleon recoil effect.
The s-wave rescattering vertex in Fig.1(b)
is customarily described with the phenomenological
Hamiltonian\cite{kr66}
\begin{equation}
{\cal H}_{1} =
4\pi \frac{\lambda_1}{m_\pi} \bar{\psi}
\mbox{\boldmath$\pi$}\!\cdot\!
\mbox{\boldmath$\pi$} \psi
+ 4\pi \frac{\lambda_2}{m^2_\pi}
\bar{\psi} \mbox{\boldmath$\tau$}\!\cdot\!
\mbox{\boldmath$\pi$} \!\times\!
\dot{\mbox{\boldmath$\pi$}} \psi
\label{eq:H1}
\end{equation}
The coupling constants
$\lambda_1$ and $\lambda_2$ determined
from the experimental pion-nucleon scattering lengths are
$\lambda_1 \sim 0.005$ and
$\lambda_2 \sim 0.05$.
Thus,  $\lambda_1 \ll \lambda_2$,
as expected from current algebra.
The calculations\cite{ms91,kr66} based on
these phenomenological vertices
yield cross sections for s-wave $\pi^0$ production
that are significantly smaller,
typically by a factor of $\sim$5,
than the experimental values\cite{meyetal90}.
There are, however,
some delicate aspects in the calculated cross section.
First, in Eq.(\ref{eq:H0}),
only the second term contributes
to $s$-wave pion production.
The suppression factor
$\sim m_\pi /m_{\mbox{\tiny N}}$
contained in this term drastically reduces
the contribution of the impulse term,
Fig.1(a),
and as a consequence the relative importance
of the two-body rescattering process, Fig.1(b), is enhanced.
However, the dominant $\lambda_2$ term
in Eq.(\ref{eq:H1})
cannot contribute to the $pp\pi^0\pi^0$ vertex
in Fig.1(b) due to its isospin structure.
Thus, a phenomenological calculation
based on Eqs.(\ref{eq:H0}) and (\ref{eq:H1})
leads to highly suppressed cross sections.

To describe the more recent theoretical developments,
it is convenient to introduce
what we call the {\it typical threshold}
({\it TT}) {\it kinematics}.
Consider Fig.1(b) in the center of mass (CM) system
with the initial and final interactions turned off.
At threshold,
$(q_0,\,\vec{q}) = (m_\pi,\,\vec{0})$,
$p_{10}'=p_{20}'=m_{\mbox{\tiny N}}$,
$\vec{p}_1^{\,\,\prime}
=\vec{p}_2^{\,\,\prime}=\vec{0}$,
so that
$p_{10}=p_{20}=m_{\mbox{\tiny N}}
+m_\pi/2$, $k_0=m_\pi/2$,
$\vec{p}_1=\vec{k}=-\vec{p}_2$ with
$|\vec{k}| = \sqrt{m_\pi m_{\mbox{\tiny N}}
+ m_\pi^2/4}$.
Needless to say, even for $\vec{q} = 0$,
the actual kinematics
for the transition process may differ
from the {\it TT kinematics}
due to the initial- and final-state interactions.
Now, for the {\it TT kinematics} we have
$k^2=-m_{\mbox{\tiny N}} m_\pi$, which implies that
the rescattering process typically probes
inter-nucleon distances $\sim$ 0.5 fm.
The reaction can therefore be sensitive
to exchange of the heavy mesons,
which play an important role
in the phenomenological meson-exchange
$N$-$N$ potentials.
Lee and Riska\cite{lr93} pointed out that
the shorter-range meson exchanges
were indeed capable of enhancing
the cross section significantly.
Meanwhile, even though
the {\em off-shell} $\pi N$ amplitudes
feature in the rescattering process
({\it e.g.,}\
$k^2=-m_{\mbox{\tiny N}} m_\pi\ne m_\pi^2$
for the {\it TT kinematics}),
there is no guarantee
that ${\cal H}_1$ of Eq.(\ref{eq:H1}) describes
the off-shell amplitudes adequately.
Hern\'{a}ndez and Oset\cite{ho95}
suggested that the $s$-wave amplitude
enhanced for off-shell kinematics
could increase the rescattering contribution
sufficiently to reproduce the experimental cross sections.
However, Ref.\cite{ho95} used
phenomenological off-shell extrapolations,
the reliability of which requires further examination.

Chiral perturbation theory
($\chi$PT)\cite{gl84,bkm95}
serves as a consistent framework
to describe the low-energy $\pi$N scattering amplitudes.
Taking advantage of this fact,
Park {\it et al.}\cite{pmmmk96}
(to be referred to as PM$^3$K)
and Cohen {\it et al.}\cite{cfmv96}
carried out the first $\chi$PT calculations for
the $pp \rightarrow pp\pi^0$ reaction.
The results of these two groups\cite{pmmmk96,cfmv96}
essentially agree with each other
on the following major points:
(1) the pion rescattering term in a $\chi$PT treatment
is significantly larger than in the conventional treatment;
(2) the sign of the rescattering term in a $\chi$PT treatment
is opposite to that obtained in the conventional approach;
(3) the enhanced rescattering term in a $\chi$PT treatment
almost cancels the impulse term,
leading to theoretical cross sections much smaller
than the observed values.
Thus a systematic treatment
of the off-shell $\pi N$ scattering amplitudes
indeed drastically changes
the $pp \rightarrow pp\pi^0$ cross section.
We note, however,
that the calculations in Refs.\cite{pmmmk96,cfmv96},
which rely on coordinate space representation,
involve potentially problematic approximations
on the kinematical variables appearing in
the energy-dependent $\pi N$ scattering amplitudes.
Namely, in deriving the r-space representation
of the two-body transition operator [see Fig. 1(b)],
one limits oneself to the Feynman amplitude
corresponding to the {\it TT kinematics}
and Fourier-transforms it with respect to
$\vec{p}_1$, $\vec{p}_2$, $\vec{p}^{\,\,\prime}_1$ and
$\vec{p}^{\,\,\prime}_2$, while keeping all the other kinematical
variables fixed at their {\it TT kinematics} values.
Although this type of simplification of kinematics
is commonly used in nuclear physics
to derive effective r-space operators,
it is expected to be much less reliable for the threshold
$pp \rightarrow pp\pi^0$ reaction.  The reason is two-fold:
First, the energy-momentum exchange due to the initial and
final-state interactions is vitally important for this process.
Secondly, the destructive interference between
the one-body and two-body terms implies that
even a small change in the two-body term can influence
the cross sections
significantly.\footnote{A detailed momentum-space
calculation by Hanhart {\it et al.}\cite{hanetal95},
with the use of the conventional Hamiltonian,
indicates the importance of lifting
the kinematical simplifications discussed here.}
In this article we carry out a $\chi$PT calculation
for the $pp \rightarrow pp\pi^0$ reaction with the use
of momentum representation,
which liberates us from the above-mentioned
kinematical simplifications.
As we shall show, this improvement
drastically changes the calculated cross section.

This paper is organized as follows.
We describe in section II
the effective Lagrangian of $\chi$PT
and a formalism we use to apply
$\chi$PT to processes involving two or more nucleons.
The details of our momentum-space calculation
are explained in section III.
In section IV we present
the numerical results and
comment on their salient features.
The final section V contains discussion
and concluding remarks.

\section{Nuclear chiral perturbation theory}

As in PM$^3$K\cite{pmmmk96},
we use the heavy-fermion formalism (HFF)
of chiral perturbation theory\cite{jm91}.
The effective Lagrangian ${\cal L}_{\rm{ch}}$
in the HFF is expanded as
\begin{equation}
{\cal L}_{\rm{ch}} = {\cal L}^{(0)} + {\cal L}^{(1)}
+ {\cal L}^{(2)}+ \cdots\;\; .
\label{eq:Lag0}
\end{equation}
${\cal L}^{(\bar{\nu})}$ represents
a term of chiral order $\bar{\nu}$
with $\bar{\nu}\equiv d + (n/2) - 2$,
where $n$ is the number of
fermion lines involved in a vertex,
and $d$ is the number of derivatives or
powers of $m_\pi$.
In order to produce the one-body and
two-body diagrams depicted
in Figs.1(a) and 1(b),
we minimally need terms with
$\bar{\nu}=0$ and $1$\cite{pmmmk96}.
Their explicit forms are\cite{bkm95}
\begin{eqnarray}
{\cal L}^{(0)} &=&
 \frac{f^2_\pi}{4} \mbox{Tr}
[ \partial_\mu U^\dagger \partial^\mu U
 + m_\pi^2 (U^\dagger +  U - 2) ]
 + \bar{N} ( i v \cdot D + g_A^{} S \cdot u ) N
\label{eq:L0}\\
{\cal L}^{(1)} &=&
-\frac{i g_A^{}}{2m_N} \bar{N}
\{ S \!\cdot\! D, v \!\cdot\! u \} N
 + 2c_1 m_\pi^2 \bar{N} N \mbox{Tr}
( U + U^\dagger - 2 )  \nonumber \\
 &&+ (c_2 \!-\! \frac{g_A^2}{8m_N}) \bar{N}
(v \!\cdot\! u)^2 N
 + c_3 \bar{N} u \!\cdot\! u  N
\label{eq:L1}
\end{eqnarray}
We have retained here only terms
of direct relevance for our present calculation.
In the above,
$U(x)$ is an SU(2) matrix
that is non-linearly related to the pion field
and that has standard chiral
transformation properties.
We use
$U(x) = \sqrt{1-[\bbox{\pi}(x)/f_\pi]^2}
+i\bbox{\tau} \!\cdot\! \bbox{\pi}(x)/f_\pi$
as in Ref. \cite{bkm95}.
The large component of the heavy-fermion field is
denoted by $N(x)$;
the four-velocity parameter $v_\mu$
is chosen to be $v_\mu=(1,0,0,0)$;
$D_\mu N$ is the covariant derivative;
$S_\mu$ is the covariant spin operator and
$u_\mu = i [\xi^\dagger \partial_\mu \xi
- \xi \partial_\mu \xi^\dagger]$, where
$\xi = \sqrt{U(x)}$ \cite{bkm95}.
The constants $c_1,c_2$ and $c_3$
have been determined from phenomenology
\cite{pmmmk96,bkm95} and their numerical values
will be discussed later.
In practical calculations, $U(x)$ is
expanded in powers of
$\bbox{\pi}(x)/f_\pi$ and
only necessary lowest order terms
of this expansion are kept.

In applying $\chi$PT to nuclei\cite{wei90},
one classifies multi-nucleon Feynman diagrams
into irreducible and reducible diagrams.
Diagrams in which every intermediate state
contains at least one meson
are categorized as irreducible diagrams,
and all others are classified as reducible diagrams.
To each irreducible diagram one assigns
the chiral order index $\nu$ defined by
$\nu = 4 - E_N - 2C + 2L + \sum_i \bar{\nu}_i$,
where $E_N$ is the number of nucleons
in the Feynman diagram,
$L$ the number of loops,
$C$ the number of disconnected parts
of the diagram, and the sum runs over all the vertices
in the Feynman graph\cite{wei90}.
It can be shown that an irreducible diagram
of order $\nu$ carries a factor
$(q/\Lambda)^\nu$, where $q$
is a generic momentum characterizing
low-energy phenomena and $\Lambda\sim 1$ GeV
is the scale parameter of $\chi$PT.
This leads to the general expectation
that the contributions of terms
with higher values of $\nu$
are significantly suppressed in the low-energy regime.
Now, the contribution of all the irreducible diagrams
(up to a specified chiral order)
is treated as an effective operator ${\cal T}$
acting on nucleonic wave functions.
The resulting nuclear matrix elements
incorporate the contributions of
the reducible diagrams.
This two-step procedure may be referred to
as {\it nuclear chiral perturbation theory}.

Applying nuclear $\chi$PT to the present case,
we write the transition amplitude
for the $pp \rightarrow pp\pi^0$ reaction as
\begin{equation}
T\,=\,\langle \Phi_f | {\cal T} | \Phi_i \rangle,
\label{eq:Tmatrix}
\end{equation}
where $|\Phi_i\rangle$ ($|\Phi_f\rangle$)
is the initial (final) two-nucleon state
distorted by the initial-state (final-state) interaction.
For formal consistency,
if ${\cal T}$ is calculated up to order $\nu,$
the nucleon-nucleon interactions that generate
$|\Phi_i\rangle$ and $|\Phi_f\rangle$
should also be calculated by summing up
all irreducible two-nucleon scattering diagrams
up to order $\nu$.
In practice, however, it is common
to use the phenomenological $N$-$N$
interactions that reproduce measured
two-nucleon observables.
This hybrid version of nuclear $\chi$PT
has been applied with great success
to electroweak transition processes\cite{pmr93},
and we shall use this phenomenological version
in this paper.

As discussed in PM$^3$K,
the lowest-order contributions
to the impulse and rescattering terms
come from the $\nu=-1$ and $\nu=1$ terms, respectively.
With the use of ${\cal L}_{\rm{ch}}$
in Eq.(\ref{eq:Lag0}),
the matrix elements
in momentum space\footnote{Here we normalize
plane-wave states according to
$<\!\vec{p}^{\,\, \prime}\!\mid \vec{p}\!>
=\delta(\vec{p}^{\,\,\prime}-\vec{p})$;
this normalization differs
from the one used in Ref.\cite{pmmmk96}.}
of these two terms are given by
\begin{eqnarray}
{\cal T}^{(-1)}
& = &\frac{i}{(2\pi)^{3/2}}
\frac{1}{\sqrt{2\omega_q}}\frac{g_A}{2f_\pi}
\sum_{j=1,2} [-\vec{\sigma}_j\cdot\vec{q}
+\frac{\omega_q}{2m_N}\vec{\sigma}_j
\cdot (\vec{p}_j + \vec{p}_j^{\,\,\prime} )]\tau_j^0,
\label{eq:Tminus1} \\
{\cal T}^{(+1)}
&=& \frac{-i}{(2\pi)^{9/2}}
\frac{1}{\sqrt{2\omega_q}}\frac{g_A}{f_\pi}
\sum_{j=1,2} \kappa(k_j,q)\frac{\vec{\sigma}_j
\cdot \vec{k}_j\tau_j^0}{k_j^2 - m_\pi^2},
\label{eq:Tplus1}
\end{eqnarray}
where $\vec{p}_j$ and $\vec{p}^{\,\,\prime}_j$ ($j=1,2$)
denote the initial and final
momenta of the $j$-th proton.
The four-momentum of the exchanged pion is defined by the
nucleon four-momenta at the $\pi NN$ vertex:
$k_j \equiv p_j-p^{\,\,\prime}_j$, where $p_j =(E_{p_j}, \vec{p}_j),
p^{\,\,\prime}_j=(E_{p^{\,\,\prime}_j}, \vec{p}^{\,\,\prime}_j)$ with
the definition $E_p=(\vec{p}^{\,\,2} + m_N^2)^{1/2}$.
The rescattering vertex function
$\kappa(k,q)$ is calculated from Eq.(\ref{eq:L1}):
\begin{equation}
\kappa(k,q)\equiv \frac{m_\pi^2}{f_\pi^2}\,
[\,2c_1 - (c_2 - \frac{g_A^2}{8m_N})
\frac{\omega_q k_0}{m_\pi^2}
 - c_3 \frac{q\cdot k}{m_\pi^2}\,]\;,
\label{eq:kappakq}
\end{equation}
where $k=(k_0,\vec{k})$
and $q=(\omega_q, \vec{q})$
represent the four-momenta
of the exchanged and final pions,
respectively. The time component of the final pion is obviously defined
by $\omega_q=(\vec{q}^2+m_\pi^2)^{1/2}$.

In what follows, we work with ${\cal T}$ defined by
\begin{equation}
{\cal T}\,=\, {\cal T}^{(-1)}+{\cal T}^{(+1)}
\equiv {\cal T}^{{\rm Imp}}
+{\cal T}^{{\rm Resc}}
\label{eq:calTtrunc}
\end{equation}
To specify the transition operator ${\cal T}$ completely,
we need the values of the low-energy coefficients
$c_1$, $c_2$ and $c_3$.
In Ref.\cite{bkm95}, these parameters
were determined from the experimental values of
the pion-nucleon $\sigma$ term,
the nucleon axial polarizability $\alpha_A$
and the isospin-even s-wave
$\pi N$ scattering length $a^+$.
The results are
\begin{equation}
c_1=-0.87\pm 0.11\;{\rm GeV}^{-1},\;\;
c_2=3.34\pm 0.27\;{\rm GeV}^{-1},\;\;
c_3=-5.25\pm 0.22\;{\rm GeV}^{-1}.
\label{eq:lecoef}
\end{equation}
As discussed in PM$^3$K,
the value of $c_2 + c_3$ can also be extracted
from the known pion-nucleon
effective range parameter $b^+$ of the
low energy pion-nucleon scattering amplitude:
\begin{equation}
b^+ = \frac{1}{2 \pi}
\left( 1 + \frac{m_{\pi}}{m_N} \right)^{-1}
\left( \frac{m_{\pi}}{f_\pi} \right)^2
( c_2 + c_3 - \frac{g_A^2}{8 m_N})
\frac{1}{{m_\pi}^2}\,.
\label{eq:bplus}
\end{equation}
Since $c_3$ in Eq. (\ref{eq:lecoef}) has been deduced
directly from $\alpha_A^{\rm exp}$,
a quantity known with a relatively high precision,
we may use the value of $c_3$
in Eq. (\ref{eq:lecoef})
to determine $c_2$ from the observed value of $b^+$.
This procedure yields
$c_2 = (4.5 \pm 0.7 )\;{\rm GeV}^{-1}$\cite{pmmmk96},
which is significantly larger
than that in Eq.(\ref{eq:lecoef}).
Thus, the errors quoted in Eq.(\ref{eq:lecoef})
do not seem to reflect the entire range of uncertainties
in the low-energy coefficients.\footnote{
In this connection, it is worth mentioning that
the reliability of the empirical value of the
$s$-wave $\pi N$ scattering length $a^+$ has
recently been questioned\cite{sigetal95}. }
It is important to examine
to what extent the existing ambiguities
in the low-energy coefficients
affect the off-shell enhancement of
the $pp \rightarrow pp\pi^0$ reaction.
In the present work,
we use as the ``standard" parameter set
the central values in Eq. (\ref{eq:lecoef})
and refer to it as the parameter set I.
We also use the parameter set II,
in which the value of $c_2$ is
changed into $c_2=4.5\;{\rm GeV}^{-1}$.
Furthermore, in developing a certain argument
later in the text, we shall be interested in
the lower end of the error bars in the empirical $c_1$.
For this we introduce the parameter set III,
which is identical to the set I except the change:
$c_1=-0.87\;{\rm GeV}^{-1}\rightarrow
c_1=(-0.87-0.11)\;{\rm GeV}^{-1}=-0.98\;{\rm GeV}^{-1}$.
Thus, we will consider three sets of low-energy coefficients:
\begin{eqnarray}
\lefteqn{{\rm Parameter}\;\;{\rm set}\;\;{\rm I}}
\nonumber\\
& &\;\;\;\;\;\;\;\;\;c_1=-0.87\;{\rm GeV}^{-1},\;\;
c_2=3.34\;{\rm GeV}^{-1},\;\;
c_3=-5.25\;{\rm GeV}^{-1}\label{eq:set1}\\
\lefteqn{{\rm Parameter}\;\;{\rm set}\;\;{\rm II}}
\nonumber\\
& &\;\;\;\;\;\;\;\;\;c_1=-0.87\;{\rm GeV}^{-1},\;\;
c_2=4.5\;{\rm GeV}^{-1},\;\;
c_3=-5.25\;{\rm GeV}^{-1}\label{eq:set2}\\
\lefteqn{{\rm Parameter}\;\;{\rm set}\;\;{\rm III}}
\nonumber\\
& &\;\;\;\;\;\;\;\;\;c_1=-0.98\;{\rm GeV}^{-1},\;\;
c_2=3.34\;{\rm GeV}^{-1},\;\;
c_3=-5.25\;{\rm GeV}^{-1}\label{eq:set3}\
\end{eqnarray}

\vspace{0.4cm}
Before embarking on numerical work
we make two remarks.
First, ${\cal T}^{(+1)}$ above
represents only the tree-diagram contribution, Fig.1(b);
loop corrections to the $\nu=-1$ impulse term
generate transition operators of order $\nu=1$.
These additional contributions generate
an effective $\pi NN$ vertex form factor
for the impulse term, Fig.1(a), but
according to PM$^3$K's estimate,
the net effect of the loop corrections after renormalization
is less than $20 \%$ of the leading-order impulse term.
We therefore neglect the loop corrections
in the present work,
even though formal consistency requires it.
We will return to this question in later work.
Secondly, the nuclear chiral counting procedure
employed above is in fact best applicable
when energy-momentum transfers
to a nucleus are small,
whereas the near-threshold
$pp \rightarrow pp\pi^0$ reaction involves
significant energy transfers $q_0\sim m_\pi$.
Hence we must exercise caution in applying
Weinberg's counting rule to this reaction.
In PM$^3$K, as far as the construction
of the transition operators was concerned,
the energy-momentum transfer due to the final pion
was ignored ({\it i.e.},
the external pion was taken to be soft,
$q_\mu\approx 0$)
in order to utilize Weinberg's original counting rule.
The physical value of $q_\mu$
was used only at the stage of calculating
the phase space integral.
The approximate nature of this method is
particularly evident for the impulse term,
where the pion-emitting nucleon is
off-shell by $\sim m_\pi$.
This off-shell nucleon must interact
at least once with the second nucleon
before losing its off-shell character.
It is then sensible to treat
an impulse diagram accompanied
by subsequent one-pion exchange
as an irreducible diagram,
even though the original Weinberg classification
would categorize it as a reducible diagram.
Cohen {\it et al.} \cite{cfmv96} proposed
a modified chiral counting rule
that takes account of this feature.
In addition, these authors argued that
the $\Delta$ degree of freedom should be
taken into account explicitly in $\chi$PT
since the $N$-$\Delta$ mass difference $\sim 2m_\pi$
is small on the chiral scale $\Lambda$
(see also Ref.\cite{jm91a}).
In the present work,
which is basically of the illustrative nature,
we do not address these issues but simply use
the transition operator given in Eq. (\ref{eq:calTtrunc})
with the view to concentrating on the pragmatic
(but in our opinion very urgent) question:
For a {\it given} version of $\chi$PT transition operators,
how important is the difference between
the conventional r-space calculation and
a p-space calculation ?

\section{Numerical calculations}
\subsection{Momentum space calculation}
It is most convenient to carry out the calculation
in the CM frame.
Referring to Fig.~1,
we denote by $\vec{p}$ and $\vec{p}^{\,\,\prime}$
the relative momenta of the two protons
before and after the pion emission, respectively.
In terms of $\vec{p}$, $\vec{p}^{\,\,\prime}$ and
the momentum of the final pion $\vec{q}$,
the momentum of each nucleon in Fig. 1 is written as
\begin{equation}
\vec{p}_1 = - \vec{p}_2 =\vec{p},\;\;\;\;\;\;
\vec{p}_1^{\,\,\prime} =
\vec{p}^{\,\,\prime} -\frac{\vec{q}}{2},\;\;\;\;\;\;
\vec{p}_2^{\,\,\prime} = -\vec{p}^{\,\,\prime}
-\frac{\vec{q}}{2}\,.
\end{equation}
With the normalization of plane-wave states
specified earlier,
the plane-wave matrix element of
the production operator defined
by Eqs.~(\ref{eq:Tminus1})-(\ref{eq:calTtrunc})
takes the following form
\begin{eqnarray}
<\vec{p}^{\,\,\prime},\vec{q}\,|\,{\cal T}\,|\,\vec{p} >\;=\;
<\vec{p}^{\,\,\prime},\vec{q}\,|\,{\cal T}^{\rm Imp}\,|\,\vec{p}>
+ <\vec{p}^{\,\,\prime},\vec{q}\,|\,{\cal T}^{\rm Resc}\,|\,\vec{p} >,
\label{eq:Trelative}
\end{eqnarray}
where
\begin{eqnarray}
<\vec{p}^{\,\,\prime},\vec{q}\,|\,{\cal T}^{\rm Imp}
\,|\,\vec{p}>
& = &\frac{i}{(2\pi)^{3/2}}\frac{1}
{\sqrt{2\omega_q}}\,\frac{g_A}{2f_\pi}
[-\vec{\sigma}_1\!\cdot\!\vec{q}
+\frac{\omega_q}{2m_N}\,\vec{\sigma}_1
\!\cdot\!(\vec{p}_1 + \vec{p}_1^{\,\,\prime})\,]
\,\delta(\vec{p}_2 - \vec{p}_2^{\,\,\prime})
\nonumber \\
& &\;\;\;+ [\,1 \leftrightarrow 2\,]
\label{eq:TImprel}
\end{eqnarray}
and
\begin{equation}
<\vec{p}^{\,\,\prime},\vec{q}\;|\,
{\cal T}^{\rm Resc}|\,\vec{p}>
\;=\; \frac{-i}{(2\pi)^{9/2}}\frac{1}{\sqrt{2\omega_q}}
\frac{g_A}{f_\pi}\frac{\kappa(k_1,q)\vec{\sigma}_1
\!\cdot\!\vec{k}_1}{k_1^2 - m_\pi^2}
\;\;+\;\; [ 1 \leftrightarrow 2\,]\,.
\label{eq:TRescrel}
\end{equation}
In the above,
the four-momentum transfer $k_j$ ($j=1,2$) is defined by 
$k_j=(E_{\vec{p}_j}-E_{\vec{p}_j^{\, \prime}},
\vec{p}_j - \vec{p}_j^{\,\,\prime})$ with 
$E_{\vec{p}_j}=( \vec{p}_j^{\,2} + m_N^2)^{1/2}$.
Since we are assuming that the nuclear states are described
by the non-relativistic Schr\"{o}dinger equation,
the $N$-$N$ potential responsible
for the initial- and final-state interactions
represents only those two-nucleon diagrams
that involve no energy transfer between the two nucleons
({\it i.e.,} three-momentum transfers only).
Hence the intermediate nucleon energies in our treatment
are given by
$E_{\vec{p}_j} = ( \vec{p}_j^{\,2} + m_N^2)^{1/2}$
at each pion-nucleon vertex.

The transition amplitude of the
$ pp \rightarrow pp\pi^0$ reaction
is evaluated by taking the matrix element
of the production operator ${\cal T}$ defined above
between the initial ($\chi^{(+)}$)
and the final ($\chi^{(-)}$)
$pp$ scattering wavefunctions.
In terms of this transition matrix element,
the total cross section is given by
\begin{eqnarray}
\sigma_{pp\rightarrow pp\pi^0}(W)
 & = & \frac{(2\pi)^4}{2v_i}\int d\vec{p}_f d\vec{q}
       \;\delta(\sqrt{4E^2_{\vec{p}_f} + \vec{q}^2} +
\omega_q - W)  \nonumber \\
  & \times &
  \frac{1}{4}\sum_{m_{s_1}m_{s_2}m_{s'_1}m_{s'_2}}
  | <\chi^{(-)}_{\vec{p}_f, m_{s'_1},m_{s'_2}},
\vec{q}\; | {\cal T} |
      \chi^{(+)}_{\vec{p}_i, m_{s_1},m_{s_2}}> |^2,
\label{eq:sigmaWa}
\end{eqnarray}
where $\vec{p}_i$ and $\vec{p}_f$ are
the {\it asymptotic} relative momenta
of the initial and final $pp$ states, respectively,
$W=2E_{\vec{p}_i} $ is the total energy,
$v_i= 2|\vec{p}_i|/E_{\vec{p}_i}$ is the asymptotic
relative velocity of the two initial protons,
and $m_{s_j}$ is the z-component of
the spin of the $j$th nucleon.
Eq.(\ref{eq:sigmaWa}) can be calculated most easily
in the partial-wave representation.
We therefore expand the transition matrix element as
\begin{eqnarray}
\lefteqn{<\!\chi^{(-)}_{\vec{p}_f,m_{s'_1},m_{s'_2}}\vec{q}
\mid{\cal T}\mid\chi^{(+)}_{\vec{p}_i,m_{s_1},m_{s_2}}\!>
\;=}\nonumber\\
& &\;\;\;\;\;\;\;\;\;\;\;\;\;\;
\sum_{S_fL_fJ_fM_f} \sum_{S_iL_iJ_iM_i}
{\cal Y}^{J_fM_f^+}_{S_fL_f}(\hat{p}_f,m_{s'_1},m_{s'_2})
{\cal Y}^{J_iM_i}_{S_iL_i}(\hat{p}_i,m_{s_1},m_{s_2})
\nonumber \\
& & \;\;\;\;\;\;\;\;\;\;\;\;\;\;\;\;
\times\sum_{l_\pi m_\pi}
Y^*_{l_{\pi}m_{\pi}}(\hat{q})
<\!p_f[L_fS_f]J_f M_f| \,{\cal T}_{l_{\pi}m_{\pi}}(q)|
\,p_i[L_iS_i]J_i M_i\!>,
\label{eq:TYY}
\end{eqnarray}
where the spin-angular function
of the anti-symmetrized two proton state is defined by
\begin{eqnarray}
{\cal Y}^{JM}_{LS}(\hat{p},m_{s_1},m_{s_2})
&=&\sum_{m_{s_1}m_{s_2}}
\frac{1-(-1)^{L+S+1}}{\sqrt{2}}
  \sum_{M_S M_L }i^L e^{i\delta_{(LS)J}}
Y^*_{L M_L}(\hat{p}) \nonumber \\
  & &\;\;\;\;\times
  <\frac{1}{2}\frac{1}{2}m_{s_1} m_{s_2}|SM_S\!>
<\!LSM_L M_S|J M\!>
\label{eq:spinangle},
\end{eqnarray}
where $\delta_{(LS)J}$ is
the NN scattering phase shift in the eigenchannel
defined by the orbital angular momentum $L$,
total spin $S$, and total angular momentum $J$.
We introduce the reduced matrix element
using the standard convention:
\begin{eqnarray}
\lefteqn{<\!p_f[L_fS_f]J_f M_f|
\,{\cal T}_{l_{\pi}m_{\pi}}(q)|
\,p_i[L_iS_i]J_i M_i\!>\;\;\equiv}\nonumber\\
& & \;\;\;\;\;\;\;\;\;\;\;\;\;\;\;\;(-1)^{J_f-M_f}
\left(
\begin{array}{ccc}
  J_f   &  l_\pi    & J_i\\
 -M_f  &  m_\pi  & M_i
\end{array}
\right)\,
<\!p_f[L_fS_f]J_f\,||\,T_{l_\pi}(q)\,||\,p_i [L_iS_i]J_i\!>\,.
\label{eq:Treduced}
\end{eqnarray}

\noindent
The substitution of
Eqs. (\ref{eq:TYY})-(\ref{eq:Treduced})
into Eq.(\ref{eq:sigmaWa}) leads to
\begin{eqnarray}
\sigma_{pp\rightarrow pp\pi^0}(W) & = &
 \frac{(2\pi)^4}{16}\frac{E_{p_i}}{p_i}
\int_{0}^{q_m}\!dq\,q^2 p_f
\sqrt{4E_{p_f}^2 + \vec{q}^{\,2}} \nonumber \\
 & \times &
 \sum_{L_i S_i J_i L_f S_f J_f l_{\pi}}
 | \frac{1}{\sqrt{4\pi}}e^{i\delta_{(L_fS_f)J_f}
+ i\delta_{(L_i,S_i)J_i}} <\!p_f[L_f S_f]J_f\,||\,
{\cal T}_{l_{\pi}}(q)||\,p_i[L_i S_i] J_i\!>|^2\nonumber\\
& & \label{eq:sigmaWb}
\end{eqnarray}
where $E_{p_f}\equiv
\{(W-\omega_q)^2 - \vec{q}^2\}^{1/2}/2$,
$p_f\equiv \sqrt{E_{p_f}^2-m_N^2}$\,,
and the maximum momentum of the pion
is given by
$q_m$ = $\sqrt{\{(W-2m_N)^2-m_\pi^2\}
\{(W+2m_N)^2-m_\pi^2\}/4W^2}$\,.
Corresponding to the decomposition
${\cal T}={\cal T}^{\rm Imp}+{\cal T}^{\rm Resc}$
in Eq. (\ref{eq:Trelative}), we have
\begin{eqnarray}
\lefteqn{<\!p_f[L_fS_f]J_f\,||\,{\cal T}_{l_\pi}(q)
\,||\,p_i [L_iS_i]J_i\!>}\nonumber\\
& &\;\;\;=<\!p_f[L_fS_f]J_f\,||\,
{\cal T}^{\rm Imp}_{l_\pi}(q)
\,||\,p_i [L_iS_i]J_i\!> +
<\!p_f[L_fS_f]J_f\,||\,{\cal T}^{\rm Resc}_{l_\pi}(q)
\,||\,p_i [L_iS_i]J_i\!>\,,\label{eq:Treldecomp}
\end{eqnarray}
where each of the reduced matrix elements
on the right-hand side
is defined similarly to Eq.~(\ref{eq:Treduced}).

For the near-threshold $\pi$ production
under consideration,
we limit ourselves to the case
where the produced pion is in s-wave ($l_\pi=0$)
and the final $pp$ state is in the $^1S_0$ state.
Accordingly, the initial $pp$ state is restricted
to be in the $^3P_0$ state.

Computation of the reduced matrix elements
in Eq. (\ref{eq:Treldecomp})
involves only the radial parts of
the scattering wavefunctions.
We calculate the momentum-space radial wavefunction
from the half-off-shell $K$-matrix
\begin{eqnarray}
R_{(LS)J,p_0}(p) = i^{-L} \cos(\delta_{(LS)J})
\left[\frac{\delta(p-p_0)}{p_0^2} +
{\cal P} \frac{K_{(LS)J}(p,p_0,W)}
{p_0^2/m_N - p^2/m_N}\right]
\label{eq:Kmatrix}
\end{eqnarray}
where ${\cal P}$ means
taking the principal-value part of the two-nucleon propagator,
and $p_0$ is the on-shell momentum defined by $W=2E_{p_0}$.
The $K$-matrix is related to the phase shift by
$\rho K_{(LS)J}(p_0,p_0,W) =
- \tan(\delta_{(LS)J})$ with $\rho=\pi p_0 m_N/2$.
By using the well-developed
numerical methods (see, {\it e.g.,} Ref.\cite{tab70}),
the $K$-matrix (or the scattering t-matrix)
can be calculated directly from the potential by solving the
Lippman-Schwinger equation in momentum space.
In terms of these radial functions,
the reduced matrix element for the impulse term
is written as
\begin{eqnarray}
\frac{1}{\sqrt{4\pi}}<\!p_f[^1S_0]\,||\,
{\cal T}^{\rm Imp}_{l_\pi}(q)
\,||\,p_i [^3P_0]\!>\; =\;
   \frac{-i}{\sqrt{(2\pi)^3 2\omega_q}}
\frac{g_A}{f_{\pi}}
\int \int\frac{d\vec{p}^{\,\,\prime}d\vec{p}}{4\pi}
  R_{^1S_0,p_f}(p') \nonumber \\
 \times \hat{p} \cdot (- \vec{q} + 
\frac{\omega_q}{m_N}\vec{p}^{\,\,\prime})
 \delta(\vec{p}^{\,\,\prime} - \vec{p} + \vec{q}/2)
  R_{^3P_0,p_i}(p)
\label{eq:TImppspace}
\end{eqnarray}
while the reduced matrix element
of the rescattering term is given by
\begin{eqnarray}
\frac{1}{\sqrt{4\pi}}<p_f[^1S_0]\,||\,
{\cal T}_{l_{\pi}=0}^{\rm Resc}(q)\,||\,p_i [^3P_0]>
 &=&  \frac{i}{\sqrt{(2\pi)^3 2\omega_q}}
\frac{2 g_A}{f_{\pi}}
\int \int\frac{d\vec{p}^{\,\,\prime}d\vec{p}}{4\pi}
  R_{^1S_0,p_f}(p') \nonumber \\
&\times& \frac{\kappa(k,q)}{(2\pi)^3}
\frac{\hat{p} \cdot \vec{k}}{ k^2 - m_{\pi}^2}
  R_{^3P_0,p_i}(p)
\label{eq:TRescpspace}
\end{eqnarray}
where $k = (k_0,\vec{k}) =
(E_{\vec{p}}-E_{\vec{p}^{\,\,\prime} -
\vec{q}/2}, \vec{p} - \vec{p}^{\,\,\prime} + \vec{q}/2)$.
As will be discussed in the next subsection,
an accurate numerical calculation of
the impulse term Eq.(\ref{eq:TImppspace})
is possible either in momentum space or in coordinate space.
By contrast, since the rescattering term
Eq.(\ref{eq:TRescpspace}) is highly nonlocal,
its accurate numerical calculation
is possible only in momentum space.

\subsection{Expressions in r-space}
We now relate our p-space calculation
to the r-space calculations in the literature.
With the use of the usual normalization
$<\!\vec{r}\,|\,\vec{r}^{\,\,\prime}\!>\,=\,
\delta(\vec{r}-\vec{r}^{\,\,\prime})$
for the r-space base vectors,
the radial functions in p- and r-space are related
to each other
by the standard Bessel transformation
\begin{eqnarray}
R_{(LS)J,p_0}(r) = \sqrt{\frac{2}{\pi}}\,i^L
\!\int\! p^2dp\,j_L(pr)R_{(LS)J,p_0}(p)\,,
\label{eq:Bessel}
\end{eqnarray}
and $R_{(LS)J,p_0}(r)$ satisfies the boundary condition
\begin{eqnarray}
R_{(LS)J,p_0}(r)
\stackrel{r\rightarrow\infty}{\longrightarrow}
\sqrt{\frac{2}{\pi}}\,[\cos(\delta_{(LS)J}) j_{L}(p_0r)
-\sin(\delta_{(LS)J})n_L(p_0r)],
\label{eq:asymptotic}
\end{eqnarray}
where $j_L(x)$ and $n_L(x)$ are the regular and
irregular spherical Bessel functions.

As for the impulse term,
the momentum-space expression,
Eq. (\ref{eq:TImppspace}), can be easily
cast into the coordinate-space expression
with the use of Eq. (\ref{eq:Bessel}).
The result is
\begin{eqnarray}
\frac{1}{\sqrt{4\pi}}<\!p_f[^1S_0]\,||\,
{\cal T}_{l_{\pi}=0}^{\rm Imp}(q)\,||\,p_i[^3P_0]\!>
   &=&  \frac{1}{\sqrt{(2\pi)^32\omega_q}}
\frac{g_A}{f_{\pi}}
   \int\!dr\, r^2  R_{^1S_0,p_f}(r) \nonumber \\
&\times&  [ ( 1 + \frac{\omega_q}{2m_N})
q j_1(qr/2) \nonumber \\
&-& \frac{\omega_q}{m_N}j_0(qr/2)
(\frac{d}{dr} + \frac{2}{r})\,]\,R_{^3P_0,p_i}(r)
\label{eq:TImprspace}
\end{eqnarray}
In the $\vec{q}\rightarrow 0$ limit,
this expression reduces to
\begin{eqnarray}
\frac{1}{\sqrt{4\pi}}<p_f[^1S_0]\,||\,
{\cal T}^{\rm Imp}_{l_\pi=0}\,||\,p_i[^3P_0]>_
{\vec{q}\rightarrow 0}
&=&   \frac{-1}{\sqrt{(2\pi)^3 2m_{\pi}}} \nonumber \\
&\times& \frac{g_Am_{\pi}}{f_{\pi}m_N}
  \int\!dr r^2  R_{^1S_0,p_f}(r)   (\frac{d}{dr} +
\frac{2}{r})  R_{^3P_0,p_i}(r)\,,
\label{eq:TImpPMK}
\end{eqnarray}
and this corresponds to Eq.~(54a) of PM$^3$K.

As stated, the rescattering operator is
highly non-local and hence its matrix element,
Eq. (\ref{eq:TRescpspace}), can be evaluated exactly
only in momentum space.
The relation between the full expression,
Eq. (\ref{eq:TRescpspace}),
and the approximate r-space form
used in the PM$^3$K calculation is as follows.
We first simplify Eq. (\ref{eq:TRescpspace})
by taking the $ \vec{q} \rightarrow 0$ limit:
\begin{eqnarray}
\frac{1}{\sqrt{4\pi}}
<\! p_f[^1S_0]\,||\,{\cal T}^{\rm Resc}_{l_\pi=0}\,||\,
p_i[^3P_0] >_{\vec{q}\rightarrow 0}
&=&  \frac{-2\pi i}{\sqrt{(2\pi)^3 2m_{\pi}}}
\frac{g_A}{f_{\pi}}
\int\! dp\,dp'\,p p'  R_{^1S_0,p_f}(p')
\frac{\bar{\kappa}}{(2\pi)^3} \nonumber \\
&\times& \left\{  2p' +  (p - x p')\log
\frac{x+1}{x-1} \right\}
R_{^3P_0,p_i}(p) \label{eq:TRescq0}
\end{eqnarray}
where
\begin{equation}
 x \equiv \frac{p^2 + p'^2 - k_0^2 + m_{\pi}^2}{2pp'},
\;\;\;\;\;\;\;\;\;\;\;
k_0  =  E_{p} - E_{p'},
\label{equx}
\end{equation}
and $\bar{\kappa}$ is defined
\ [{\it cf.} Eq.(\ref{eq:kappakq})] as
\begin{equation}
\bar{\kappa}  =
\frac{m_{\pi}^2}{f_{\pi}^2}\,\left[\,2c_1 -
(c_2 -\frac{g_A^2}{8m_N} + c_3)
\frac{k_0}{m_{\pi}}\right]
\label{equkap}
\end{equation}
If we furthermore freeze $k_0$, the energy variable
of the exchanged pion, at the $fixed$ value
$k_0=m_\pi/2$ corresponding to the threshold pion
production,
then we are back with the {\it TT kinematics}
explained in section I.
For the {\it TT kinematics},
\begin{equation}
x \;\stackrel{k_0=m_\pi/2}{\longrightarrow}\;\;
 \frac{p^2 + p'^2 + 3m_{\pi}^2/4}{2pp'},
\label{eq:xPMK}
\end{equation}
and $\bar{\kappa}$ becomes a constant,
which was denoted by $\kappa_{th}$
in Eq. (32) of PM$^3$K:
\begin{equation}
\bar{\kappa}\;\stackrel{k_0=m_\pi/2}
{\longrightarrow}\;\;
 \kappa_{th}\equiv
\frac{m_{\pi}^2}{f_{\pi}^2}\,\left[
\,2c_1 - \frac{1}{2}(c_2 -\frac{g_A^2}{8m_N})
-\frac{1}{2}c_3\right]
\label{eq:kappabarth}
\end{equation}
We refer (as in PM$^3$K) to the simplifications
Eqs. (\ref{eq:TRescq0}), (\ref{eq:xPMK}),
(\ref{eq:kappabarth})
as the {\it fixed kinematics approximation}.
In the {\it fixed kinematics approximation},
the right-hand side of Eq. (\ref{eq:TRescpspace})
becomes sufficiently simple to be recast
into an r-space expression via the transformation
Eq.(\ref{eq:Bessel}).
The resulting expression agrees with the r-space form
Eq.(54b) of PM$^3$K.

\medskip
\section{Numerical Results}

To calculate the reduced matrix element
$<\!p_f[^1S_0]\,||\,
{\cal T}^{\rm Imp}_{l_\pi}(q)
\,||\,p_i [^3P_0]\!>$,
we can use either Eq.(\ref{eq:TImppspace})
or Eq.(\ref{eq:TImprspace});
these are completely equivalent to each other.
For numerical computations, however,
we find the latter more convenient.
On the other hand, the reduced matrix element
for the rescattering term,
$<p_f[^1S_0]\,||\,
{\cal T}_{l_{\pi}=0}^{\rm Resc}(q)\,||\,p_i [^3P_0]>$,
is calculated with the use of
the p-space expression, Eq.(\ref{eq:TRescpspace}).
Calculations that are based on the full expressions,
Eq.~(\ref{eq:TImppspace})(or Eq.~(\ref{eq:TImprspace}))
and Eq.~(\ref{eq:TRescpspace})
are referred to as {\it full p-space}
calculations.
For the sake of comparison,
we also carry out calculations
with the use of the approximate expressions
discussed in the preceding section.
For given reduced matrix elements,
the total cross section is obtained
via Eqs.~(\ref{eq:sigmaWb}), (\ref{eq:Treldecomp}).

The radial functions for the $pp$ scattering states
that appear in Eqs.~(\ref{eq:TRescpspace}),
(\ref{eq:TImprspace}) are to be generated
with the use of realistic nucleon-nucleon interactions.
We adopt here as our standard choice
the Argonne V18 potential\cite{wir95}.
For the purpose of comparison,
we also carry out supplementary calculations
with the Reid soft-core potential\cite{reid}.
As regards the low-energy coefficients,
$c_1$, $c_2$ and $c_3$,
our standard choice is
the parameter set~I [Eq. (\ref{eq:set1})],
and we use it throughout
unless otherwise stated.

We note here that the accuracy of the numerical integration for the
rescattering term, Eq. (\ref{eq:TRescpspace}),
can be achieved only when the high momentum component of
$pp$ wavefunction is accounted for correctly. This is due
to the fact that the $c_2$ and $c_3$ terms in $\kappa (k,q)$,
Eq.(\ref{eq:kappakq}),
grow linearly with the momentum for higher momenta.
We have found that
a very large number of mesh points are needed
to achieve sufficient numerical accuracies
in evaluating the rescattering term.

Figs.~2 - 4 show the total cross sections
calculated with the Argonne V18 potential
and with the parameter set I.
It is informative to first
consider the behavior of the individual contributions
of the impulse and rescattering terms
and then proceed to discuss the behavior of
their coherent sum.
Figs.~2 and 3 give the individual contributions
of the impulse and rescattering terms, respectively,
while Fig.~4 presents the cross sections
due to the coherent sum of these two terms.

The cross sections in Fig.~2 have been obtained
by retaining only the impulse term
in the decomposition Eq. (\ref{eq:Treldecomp}).
The solid curve corresponds
to the full  calculation
based on Eq. (\ref{eq:TImprspace}),
while the broken line has been obtained
with the $\vec{q}\rightarrow 0$ approximation,
Eq. (\ref{eq:TImpPMK}).  Our numerical results
for the latter case coincide with those in PM$^3$K.
As expected,
the $\vec{q}=0$ approximation employed in PM$^3$K
becomes less accurate as the collision energy increases.
Although the difference between the two results
is not extremely large as far as the isolated contribution
of the impulse term is concerned,
yet we should remember
that, depending on the pattern of interference
between the impulse and rescattering terms,
even a rather minor change in the impulse contribution
may affect the cross section significantly.
It is therefore recommended
to avoid the $\vec{q}=0$ approximation
in computing the impulse term.

We next discuss the contribution
of the rescattering term.
Fig.~3 shows the cross sections obtained
by retaining only the rescattering term
in Eq. (\ref{eq:Treldecomp}).
The solid curve corresponds to
the full p-space calculation.
The dash-dotted curve gives the result
obtained with the use of
the approximation Eq.~(\ref{eq:TRescq0}),
in which $\vec{q}$ is fixed to be zero
while the energy-transfer $k_0$
in the exchanged pion propagator and in
the coefficient $\kappa(k,q)$ is treated exactly.
The dashed curve represents the results
of the {\it fixed kinematics approximation},
viz., $\vec{q}\equiv 0$ and
$k_0 \equiv m_{\pi}/2$;
our numerical results for this case reproduce
those obtained in PM$^3$K.
The difference between the dash-dotted and
dashed lines clearly indicates
that the non-local effects due to the $k_0$-dependence
enhance the cross sections by a very large factor.
The smaller but still substantial difference
between the solid and dash-dotted curve
implies that the $\vec{q}$-dependence of
the pion propagator and of the rescattering
coupling coefficient $\kappa$
cannot be neglected in a quantitative calculation
of the $pp \rightarrow pp\pi^0$ cross section.
The solid curve and the dashed curve differ
by a factor as large as $\sim 10$.
Thus the magnitude of the rescattering amplitude obtained
in the full p-space calculation
is $\sim 3$ times larger than that obtained
in the {\it fixed kinematics approximation}.
In Refs.\cite{pmmmk96,cfmv96},
the impulse and rescattering terms gave
transition amplitudes of about the same magnitude,
and this feature is also visible
in Figs.~2 and 3 (compare the dashed lines therein).
In the full p-space calculation, however,
it is the rescattering term that is dominant, and
this feature drastically changes the interference pattern
of the two terms, as discussed immediately below.

In all the cases shown in Figs.~2 and 3,
our numerical results indicate that the reduced matrix element
for the impulse term
$<\!p_f[^1S_0]\,||\,
{\cal T}^{\rm Imp}_{l_\pi}(q)
\,||\,p_i [^3P_0]\!>$ and
that for the rescattering term
$<p_f[^1S_0]\,||\,
{\cal T}_{l_{\pi}=0}^{\rm Resc}(q)\,||\,p_i [^3P_0]>$
have opposite signs so that they interfere
with each other destructively.
The cross sections given by the coherent sum
of these two terms are given in Fig.~4.
The solid curve represents $\sigma_{\rm full}$,
the cross section obtained
in the full p-space calculation.
The dash-dotted curve corresponds to
the approximation Eq.~(\ref{eq:TRescq0}).
The cross section $\sigma_{\rm fix.\,kin.}$
corresponding to the {\it fixed kinematics approximation},
if plotted in the same scale,
would be hardly visible.
The impulse and rescattering contributions
in this case cancel each other almost perfectly,
leading to an extremely small value of
$\sigma_{\rm fix.\,kin.}$.
This is consistent with the previous r-space
calculations\cite{pmmmk96,cfmv96}.
To illustrate the calculated $\sigma_{\rm fix.\,kin.}$,
we multiply it
by a factor of 30 to obain the dashed curve in Fig.4.
When the approximation $k_0=m_\pi/2$ is removed,
the scattering term is substantially enhanced
(see Fig.~2),
and as a consequence there is only a partial cancellation
between the impulse and rescattering terms.
The corresponding cross section (dash-dotted curve)
therefore becomes significantly larger than
$\sigma_{\rm fix.\,kin.}$.
This tendency is even more prominent
as we go to the full p-space calculation
by removing the $\vec{q}\rightarrow 0$ approximation;
the solid line representing $\sigma_{\rm full}$
exhibits a huge enhancement over
$\sigma_{\rm fix.\,kin.}$.

The results in Fig.~4 indicate
that the extremely small cross sections
reported in Refs.\cite{pmmmk96,cfmv96}
are an accidental consequence
of the kinematical approximations used there
rather than a general feature
of calculations based on nuclear $\chi PT$.
It is to be noted, however, that
$\sigma_{\rm full}$ in the present calculation
is still significantly smaller
than the observed cross sections.
Our calculation here does {\it not} include
the Coulomb repulsion effect
in the initial and final $pp$ states.
The inclusion of the Coulomb effect is expected to further
reduce the cross section\cite{ms91},
worsening the discrepancy between
the calculated and observed cross sections.

We have already mentioned that
the results presented in Figs. 2-4 are obtained
with the Argonne V18 potential
and with the parameter set I, Eq.(\ref{eq:set1}).
To provide some measure of the sensitivity
of the cross section to input $NN$ potentials,
we have repeated a full p-space calculation
with the use of the Reid soft-core potential\cite{reid},
while retaining the parameter set I
for the low-energy coefficients.
The resulting cross section is given in Fig.~5
(dashed curve) along with the result
for the Argonne V18 potential (solid curve).
The difference between these two results
is not significant.

To examine the sensitivity
to different choices of the low-energy coefficients,
we have carried out full p-space calculations,
with the Argonne V18 potential,
for the parameter set II,
Eq.~(\ref{eq:set2}), and for the parameter set III,
Eq.~(\ref{eq:set3}).
The dashed curve in Fig.~6
represents the cross section corresponding to
the parameter set II.
This curve is almost indistinguishable
from the solid curve obtained
with the parameter set I
even though these two sets
have significantly different values of $c_2$.
At first sight, this feature is puzzling because
the large difference between
$\sigma_{\rm full}$ and
$\sigma_{\rm fix.\,\,kin.}$ seen in Fig.~4
seems to suggest that the cross sections
should also be sensitive to the parameters
contained in the vertex function $\kappa(k,q)$.
On closer examination, however,
the lack of sensitivity to $c_2$
in the full p-space calculation has a simple explanation.
In the near-threshold region ($\vec{q} \rightarrow 0$) 
under consideration,
the $c_2$- and $c_3$- terms in the vertex function
$\kappa (k,q)$, Eq.~(\ref{eq:kappakq}),
depend on a factor $k_0$ = $E_{p} - E_{p^\prime}$
\ [see also Eq.~(\ref{equkap})].
This energy factor
introduces a zero in the transition operator
at $|\vec{p}|$ = $|\vec{p}^{\;\prime}|$,
which drastically 
reduces the contributions from the $c_2$- and $c_3$-terms to
the integration over the scattering wavefunctions.
As a result, the rescattering term is 
dominated by the $c_1$-term
in the full p-space calculation.
In contrast, in the {\it fixed kinematics approximation}(see Eq.(37)),
the transition operator arising from the $c_2$- and
$c_3$-terms are momentum-independent and
of comparable magnitude but of
opposite sign to the transition operator coming from the $c_1-$term.
This means a substantial cancellation between these two operators
at all momenta and hence the rescattering transition matrix
element can be a sensitive function of 
the low energy coefficients
including $c_2$ in the {\it fixed kinematics approximation} calculation.
This feature originating from the treatment
of $k_0$ essentially accounts for
the large difference (by a factor of $\sim 3$),
discussed earlier, between the rescattering amplitudes
obtained in the full p-space calculation and
in the {\it fixed kinematics approximation}
(see the discussion on Fig.~3).
The same feature also explains why
the cross section in the full treatment
is hardly sensitive to $c_2$
(and $c_3$).

The dominance of the $c_1$-term
in the full p-space calculations implies that
the cross section should be sensitive to
an input value of $c_1$,
a coefficient linked to the
pion-nucleon sigma term \cite{bkm95}.
In Fig.~6 we compare the results of
full p-space calculations
corresponding to the parameter sets I and III,
which differ only in the value of $c_1$.
The cross sections indeed vary substantially
as we change the value of $c_1$
from the currently accepted central value
to one end of the error bars [see Eq.~(\ref{eq:lecoef})].
We remark, however, that even with the most
favorable choice of $c_1$ the calculated cross sections
are significantly smaller than the observed values.

\section{Discussion and Conclusions}

Our calculations in this article are based
on the transition operators derived
from a particular version of the chiral effective Lagrangian,
the one used in PM$^3$K\cite{pmmmk96}.
As mentioned in section II,
other choices of the effective Lagrangian are possible,
although the main features of our results may be
stable against these changes.
This point requires further examinations.
A related issue is the choice of an explicit form
of $U(x)$ as a function of the pion fields $\bbox{\pi}(x)$.
Needless to say, if we calculate everything exactly,
the results should be independent of particular choices.
However, in any practical calculations
one needs to expand $U(x)$ in powers of $\bbox{\pi}(x)$
and truncate the series, which nullifies the formal
equivalence of various choices.
Thus, in an approximate calculation such as the present one,
the $pp \rightarrow pp\pi^0$ transition amplitude
does depend on the off-shell $\pi N$ amplitude,
which varies when different forms for $U(x)$ are used.
We have employed here a particular form for $U(x)$,
and the stability of our results against
other choices is yet to be studied.

In our calculation,
the rescattering transition matrix element
corresponding to the $c_2$- and $c_3$-terms are
found to be rather sensitive
to the higher momentum components of
the nuclear ($p p$) wave function;
as a matter of fact, the relevant range of momentum
is not much smaller than the chiral scale $\Lambda$.
This uncomfortable feature is in fact shared also by
other known applications of nuclear $\chi$PT,
and a satisfactory solution to this problem
seems to require the study of terms
with higher chiral orders than considered here.

Keeping in mind all the caveats stated above,
we still consider it almost certain that,
in any reasonably realistic $\chi$PT calculations,
the rescattering term dominates over the impulse term
and their signs are opposite to each other.
This implies that the heavy-meson
exchange contributions
considered in Ref.\cite{lr93}
cannot be invoked as a possible mechanism
to enhance $\sigma_{\rm full}$ to bring it closer
to the observed cross sections.
Since it is established that these
heavy-meson contributions
have the same sign as the impulse term,
the addition of these extra contributions
to the transition amplitude obtained in
the full p-space calculation would
result in a destructive interference.
Thus the heavy-meson contributions such as considered
in Ref.\cite{lr93} suppresses the cross section
instead of enhancing it.
Most recently, van Kolck, Miller and Riska\cite{kmr96}
considered a yet another diagram involving
$\rho-\omega$ exchanges.
Since the sign of this additional contribution is unknown,
one way to bring our $\sigma_{\rm full}$
closer to the experimental cross sections
is to include this new contribution
assigning to it the favorable sign.
Needless to say, from a $\chi$PT point of view,
this is a highly {\it ad hoc} prescription,
but it is a possibility.

As mentioned in section II, we have not considered here
the loop corrections to the impulse term,
corrections that lead to the form factor effect
for the impulse vertex.
Qualitatively speaking, the inclusion of this effect
is expected to reduce the contribution of
the impulse term, further enhancing the dominance
of the rescattering term.
This aspect awaits further study.

We now summarize.
For the $pp \rightarrow pp\pi^0$ reaction near threshold,
we have calculated the cross sections,
using the effective transition operators
derived from chiral perturbation theory ($\chi$PT)
and employing the {\it momentum-space representation}.
Our p-space calculation is free from
the various kinematical approximations
that went into the previous
$\chi$PT calculations\cite{pmmmk96,cfmv96}
based on the r-space representation.
In what we call {\it full} p-space calculations,
we have retained all the energy-momentum dependence
in the transition vertices as dictated by $\chi$PT.
The results of the previous r-space calculations
are recovered by introducing into the present
full p-space calculation
the {\it fixed kinematics approximation},
which consists in freezing
the momentum $\vec{q}$ of the final pion
and the energy variable $k_0$ of the exchanged pion
at their values corresponding to the threshold kinematics.
The results of the full p-space calculation
indicate that the {\it fixed kinematics approximation}
is not justified.
Specifically, we have found
that in the full p-space calculation
the contribution of the rescattering term
\ [Fig. 1(b)] is enhanced by a factor of $\sim 3$
over the value obtained in the approximate r-space
calculations\cite{pmmmk96,cfmv96}.
As a result, the rescattering term
now dominates over the impulse (Born) term,
removing the near complete cancellation
between these two terms
that existed in the previous calculations.
This means that the extremely small cross section
reported in Ref.\cite{pmmmk96,cfmv96}
becomes much larger
in the full p-space calculation [see Fig.4].
The enhancement, however, is not sufficient
to explain the observed cross sections.
As our work is based on a particular version
of the chiral Lagrangian used in Ref.\cite{pmmmk96},
it remains to be seen to what extent
the use of other chiral Lagrangians would affect the results.
It is also interesting to study
whether the remaining discrepancy between
the observed and calculated cross sections
can be explained in terms of diagrams involving
heavy meson exchanges\cite{kmr96}.
Thus, the near-threshold
$pp \rightarrow pp\pi^0$ reaction
invites further detailed investigations.
The principal lesson we draw from the present calculation
is that in any serious $\chi$PT calculations
one must use what is called here the full p-space calculation,
avoiding the conventional approximations
employed in r-space calculations..

\acknowledgements

This work is supported in part
by the National Science Foundation,
Grant No. PHYS-9602000, by
the U.S. Department of Energy,
Nuclear Physics Division,
Contract No. W-31-109-ENG-38,
and by the Grant-in-Aid of
Scientific Research, the Ministry of Education,
Science and Culture, Japan,
Contract No.07640405.

\newpage
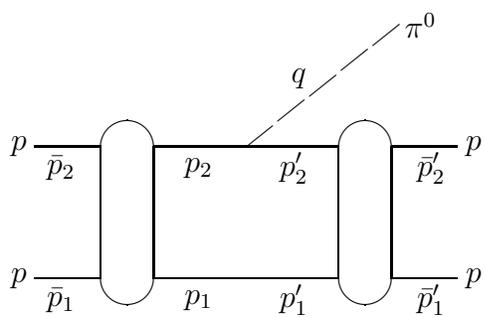
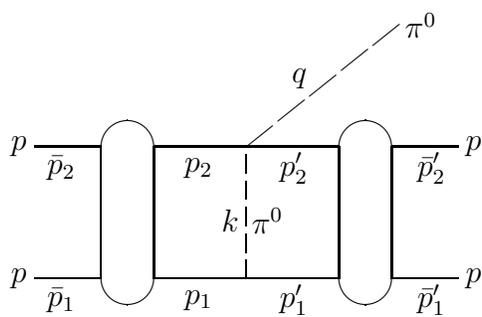
\begin{figure}
\caption[]{Diagrams contributing
to $pp\rightarrow pp\pi^0$:
impulse term (a) and rescattering term (b).
Besides the four-momenta specified
in the figures, we use in the text the following notations:
$\vec{p}_i \equiv$ relative momentum between
$\bar{\bf{p}}_1$ and $\bar{\bf{p}}_1$;
$\vec{p}_f \equiv$ relative momentum between
$\bar{\bf{p}}^{\prime}_1$
and $\bar{\bf{p}}^{\prime}_2$;
$\vec{p}\equiv$ relative momentum between
$\bf{p}_1$ and $\bf{p}_2$.
$\vec{p}^{\prime}\equiv$ relative momentum between
$\bf{p}^{\prime}_1$ and $\bf{p}^{\prime}_2$;
$W\equiv\bar{p}_{10}+\bar{p}_{20}$ =
total energy.}
\end{figure}

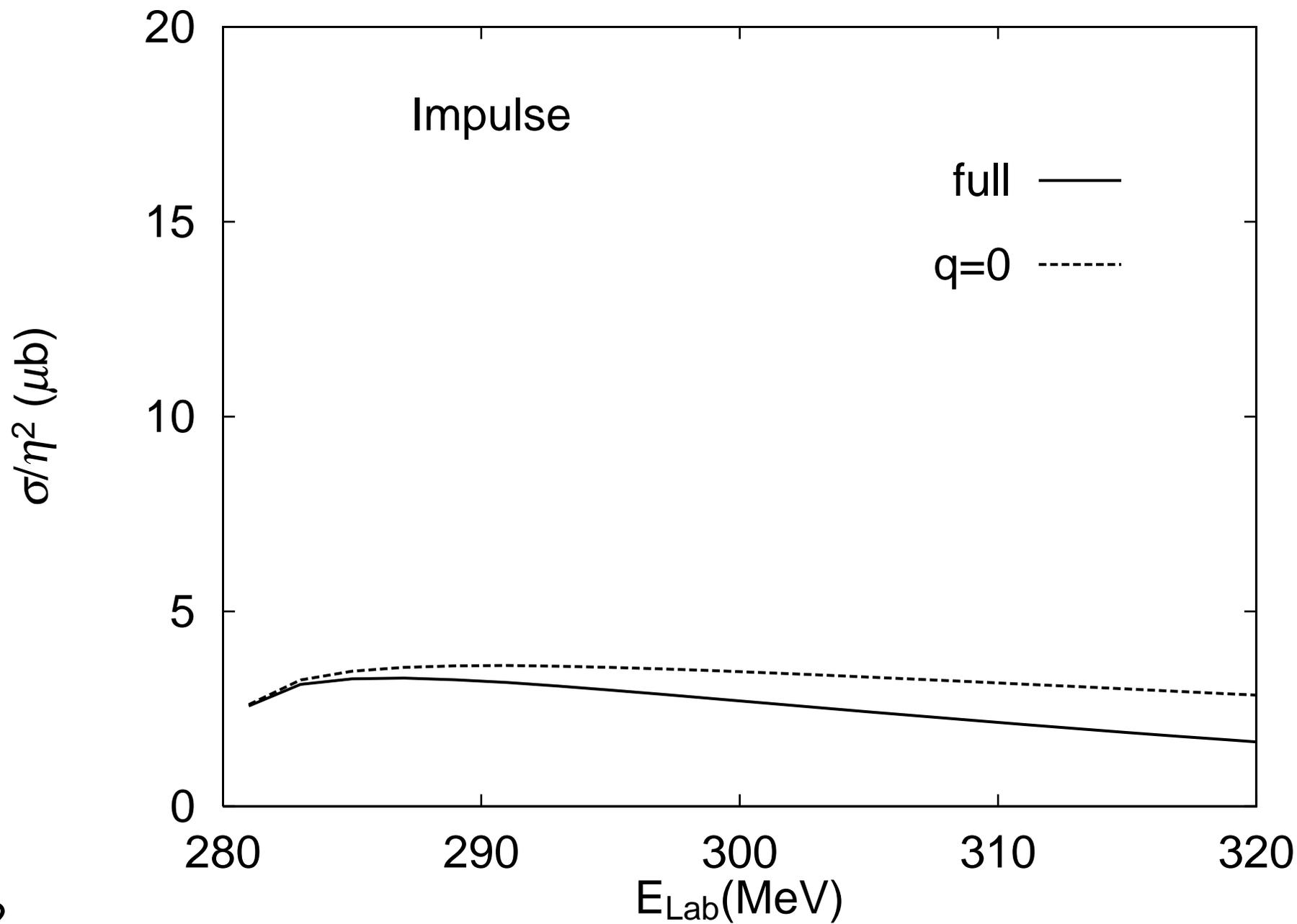
\begin{figure}
\caption[]{Total cross sections calculated
with the impulse term alone.
Solid curve: full p-space calculation;
dashed curve: $\vec{q}=0$ approximation.}
\end{figure}

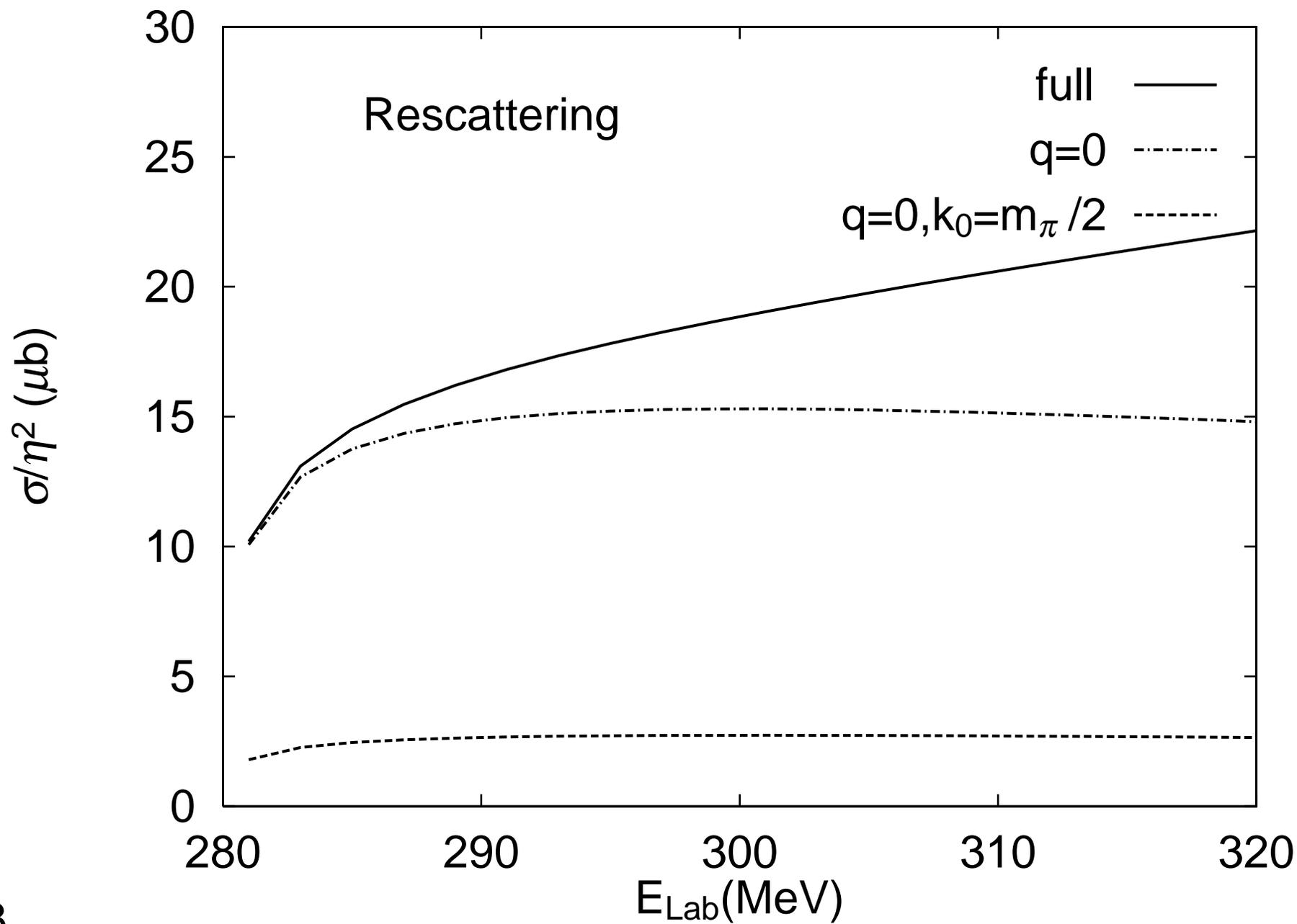
\begin{figure}
\caption[]{Total cross sections calculated
with the rescattering amplitude alone.
Solid curve: full p-space calculation;
dash-dotted curve: $\vec{q}=0$ approximation;
dashes curve: {\it fixed kinematics approximation}. }
\end{figure}

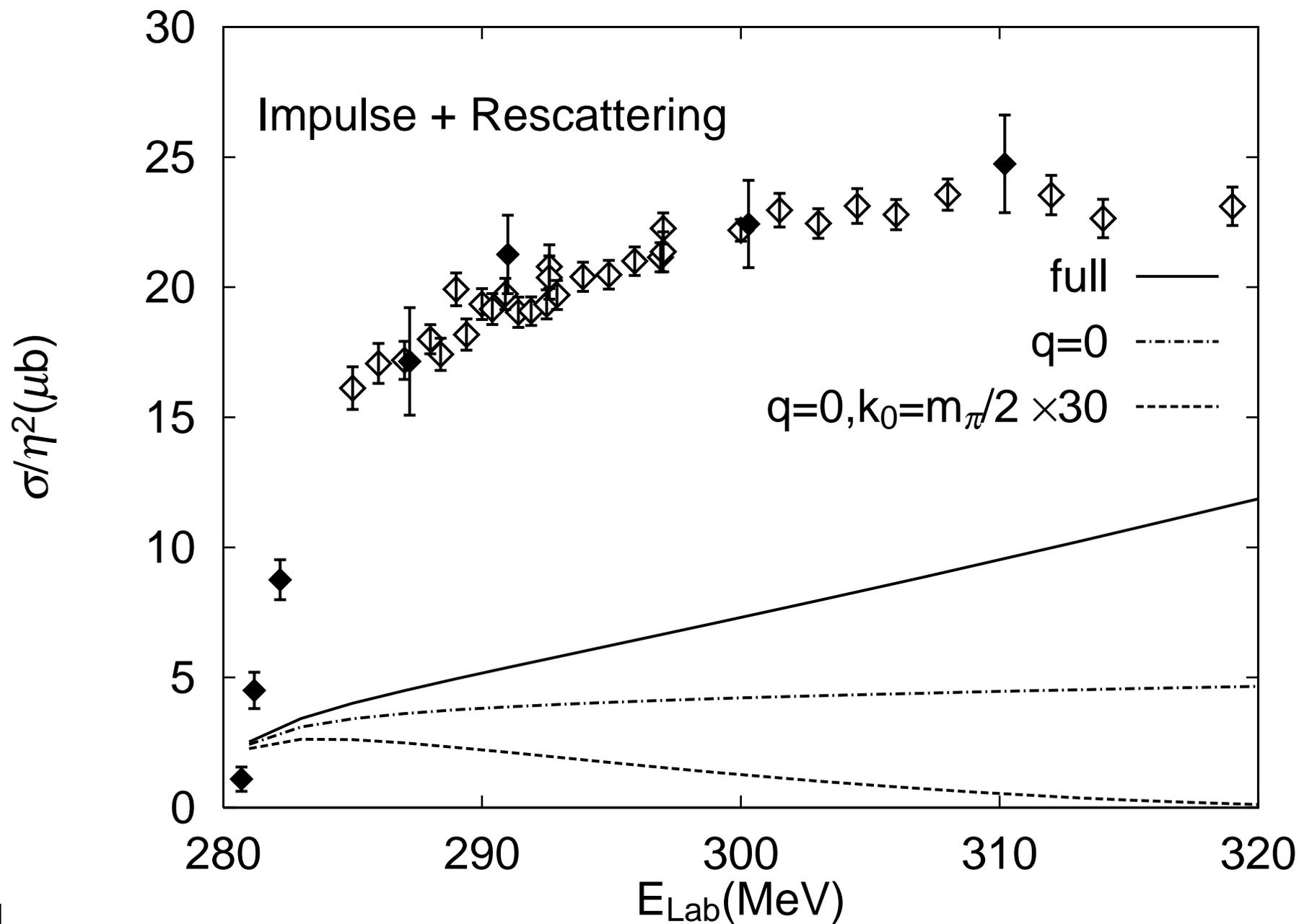
\begin{figure}
\caption[]{Total cross sections that
include both impulse and rescattering amplitudes.
Solid curve: full p-space calculation;
dash-dotted curve: $\vec{q}=0$ approximation;
dashed curve: $\sigma_{\rm fix.\,\,kin.}\times 30$.
The experimental points are
from Ref.\cite{meyetal90} (open diamonds)
and Ref.\cite{Uppsala} (solid diamonds).}

\end{figure}

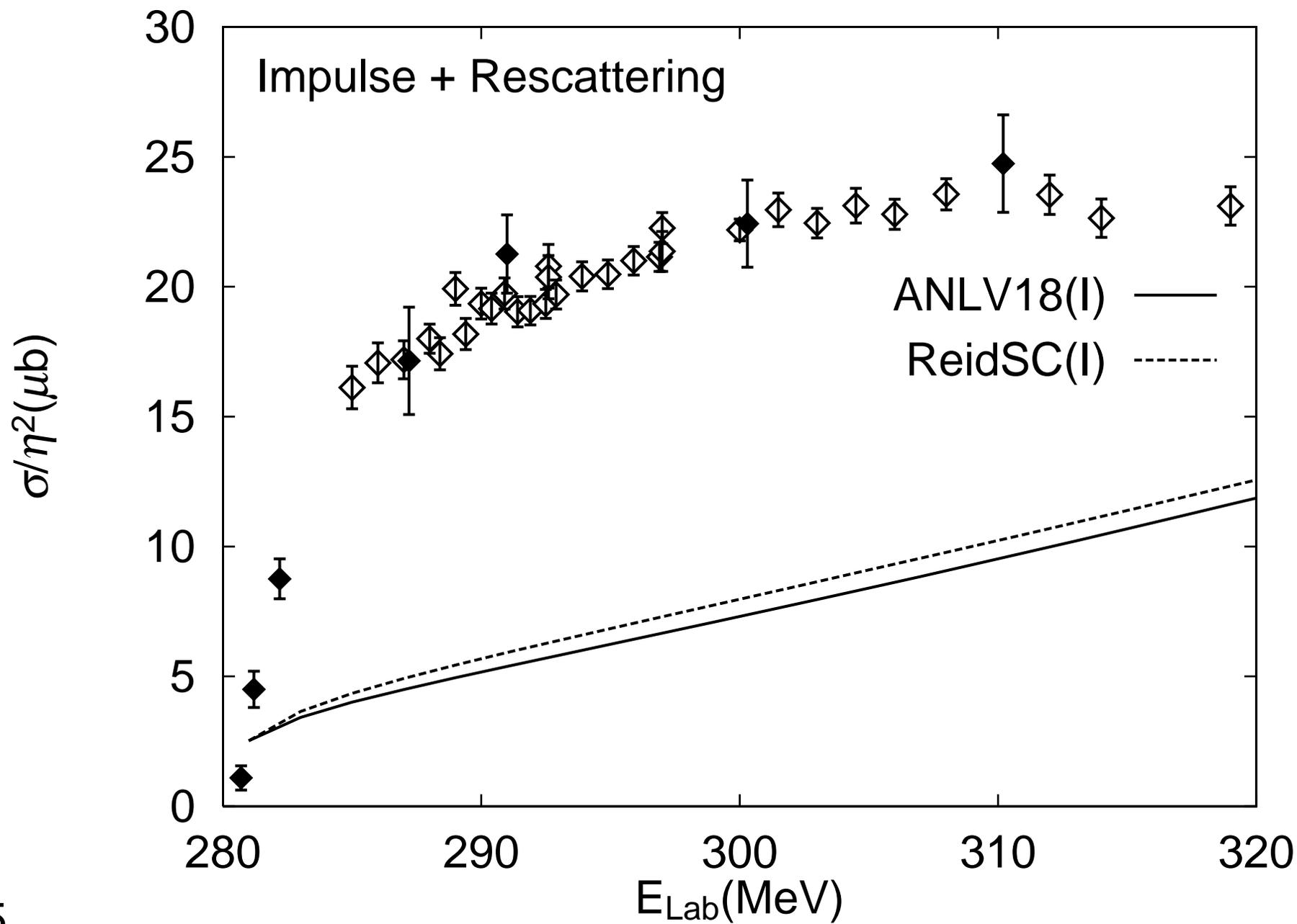
\begin{figure}
\caption[]{Total cross sections obtained
in full p-space calculations
with the Argonne V18 potential (solid curve),
and with the Reid soft-core potential (dashed curve).
The experimental points are
from Ref.\cite{meyetal90} (open diamonds)
and Ref.\cite{Uppsala} (solid diamonds).}

\end{figure}

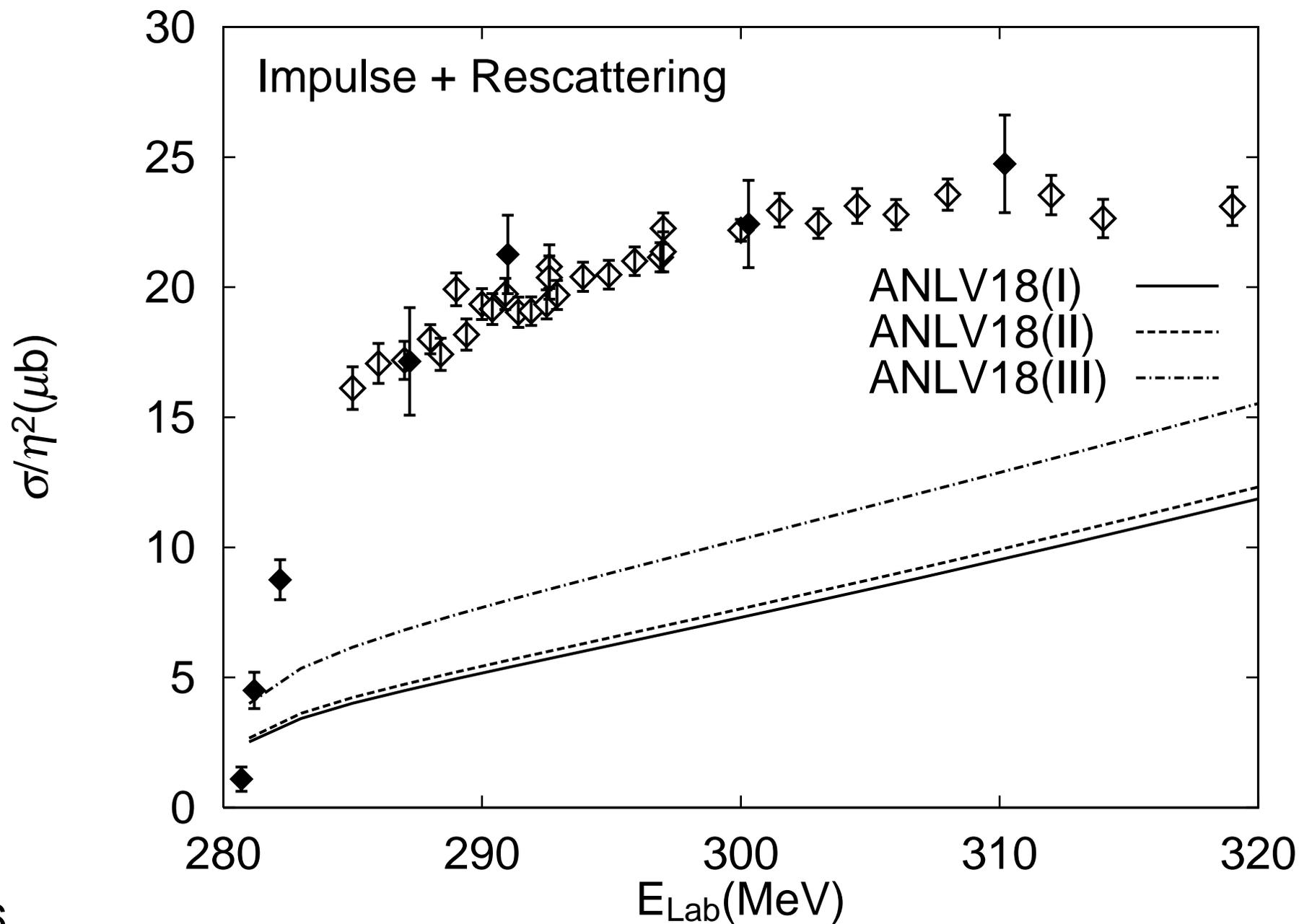
\begin{figure}
\caption[]{Total cross sections obtained
in full p-space calculations
with the parameter set I (solid curve),
the parameter set II (dashed curve)
and the parameter set III (dash-dotted curve).
The experimental points are
from Ref.\cite{meyetal90} (open diamonds)
and Ref.\cite{Uppsala} (solid diamonds).}

\end{figure}

\newpage
\begin{picture}(400,155)


\put(0,40){\line(1,0){25}}
\put(45,40){\line(1,0){70}}
\put(135,40){\line(1,0){25}}

\put(0,90){\line(1,0){25}}
\put(45,90){\line(1,0){70}}
\put(135,90){\line(1,0){25}}

\put(80,90){\line(5,4){10}}
\put(92,99.6){\line(5,4){10}}
\put(104,109.2){\line(5,4){10}}
\put(116,118.8){\line(5,4){10}}
\put(128,128.4){\line(5,4){10}}

\put(35,65){\oval(20,70)}
\put(125,65){\oval(20,70)}

\put(62,82){\makebox(0,0){$p_2$}}
\put(98,82){\makebox(0,0){$p_2'$}}
\put(62,32){\makebox(0,0){$p_1$}}
\put(98,32){\makebox(0,0){$p_1'$}}
\put(100,116){\makebox(0,0){$q$}}

\put(146,136){\makebox(0,0){$\pi^0$}}
\put(-6,90){\makebox(0,0){$p$}}
\put(166,90){\makebox(0,0){$p$}}
\put(-6,40){\makebox(0,0){$p$}}
\put(166,40){\makebox(0,0){$p$}}

\put(10,82){\makebox(0,0){$\bar{p}_2$}}
\put(150,82){\makebox(0,0){$\bar{p}_2'$}}
\put(10,32){\makebox(0,0){$\bar{p}_1$}}
\put(150,32){\makebox(0,0){$\bar{p}_1'$}}

\put(66,0){\makebox(0,0){fig. 1(a)}}



\put(210,40){\line(1,0){25}}
\put(255,40){\line(1,0){70}}
\put(345,40){\line(1,0){25}}

\put(210,90){\line(1,0){25}}
\put(255,90){\line(1,0){70}}
\put(345,90){\line(1,0){25}}

\put(290,90){\line(5,4){10}}
\put(302,99.6){\line(5,4){10}}
\put(314,109.2){\line(5,4){10}}
\put(326,118.8){\line(5,4){10}}
\put(338,128.4){\line(5,4){10}}

\put(290,40){\line(0,1){7}}
\put(290,50){\line(0,1){7}}
\put(290,60){\line(0,1){7}}
\put(290,70){\line(0,1){7}}
\put(290,80){\line(0,1){7}}

\put(245,65){\oval(20,70)}
\put(335,65){\oval(20,70)}

\put(272,82){\makebox(0,0){$p_2$}}
\put(308,82){\makebox(0,0){$p_2'$}}
\put(272,32){\makebox(0,0){$p_1$}}
\put(308,32){\makebox(0,0){$p_1'$}}
\put(310,116){\makebox(0,0){$q$}}

\put(356,136){\makebox(0,0){$\pi^0$}}
\put(204,90){\makebox(0,0){$p$}}
\put(376,90){\makebox(0,0){$p$}}
\put(204,40){\makebox(0,0){$p$}}
\put(376,40){\makebox(0,0){$p$}}
\put(298,62){\makebox(0,0){$\pi^0$}}

\put(220,82){\makebox(0,0){$\bar{p}_2$}}
\put(360,82){\makebox(0,0){$\bar{p}_2'$}}

\put(220,32){\makebox(0,0){$\bar{p}_1$}}
\put(360,32){\makebox(0,0){$\bar{p}_1'$}}
\put(284,62){\makebox(0,0){$k$}}

\put(276,0){\makebox(0,0){fig. 1(b)}}

\end{picture}
\end{document}